# Shack–Hartmann wavefront sensor with array of phase coded masks

NITIN DUBEY,* RAVI KUMAR, JOSEPH ROSEN

*School of Electrical and Computer Engineering, Ben-Gurion University of the Negev, P.O. Box 653, Beer-Sheva 8410501, Israel*
*Corresponding author: nitinyog@post.bgu.ac.il*

**Shack-Hartmann wavefront sensors (SHWS) are generally used to measure the wavefront curvature of light beams. Measurement accuracy and the sensitivity of these sensors are important factors for better wavefront sensing. In this study, we demonstrate a new type of SHWS with better measurement accuracy than the regular SHWS. The lenslet array in the regular SHWS is replaced with an array of coded phase masks and the principle of coded aperture correlation holography (COACH) is used for wavefront reconstruction. Sharper correlation peaks achieved by COACH improve the accuracy of the estimated local slopes of the measured wavefront and consequently improve the reconstruction accuracy of the overall wavefront. Experimental results confirm that the proposed method provides a lower mean square wavefront error by one order of magnitude in comparison to the regular SHWS.**

Shack-Hartmann wavefront sensor (SHWS) was developed for optical metrology purposes [1] and later has been used in various other fields, such as adaptive optics [2], microscopy [3], retinal imaging [4], and high-power laser systems [5]. Usually, in SHWS a microlens array is used to measure the wavefront local gradients. For this, the incident wavefront is spatially sampled such that each lenslet focuses the local sub-aperture into an imaging sensor located at the focal plane of the lenslet. The position of the focused spot corresponds to the average slope of the local wavefront of each microlens. The local slope of the wavefront can be evaluated by calculating the spot displacement from the optical axis. Afterward, a wavefront reconstruction algorithm is utilized to estimate the complete wavefront using these estimated slopes [6].

Recently, an incoherent digital holography technique called coded aperture correlation holography (COACH) [7,8] has been developed. In this study, we present a new technique of COACH-based SHWS that increases the accuracy of wavefront sensing. Instead of using the regular lenslet array, a coded phase mask (CPM) is introduced in each sub-aperture of the array. Based on several recently published studies [9–11] about interferenceless COACH, we conclude that using a certain CPM can yield at the end of the process a spot that is narrower than the focused spot of the regular SHWS. In any SHWS, the size of each spot constrains the sensitivity and accuracy of the measured slope of the local wavefront. Thus, by use of the CPM array, one can achieve narrower correlation peaks which improve the accuracy of the measured wavefront.

Modified Gerchberg-Saxton algorithm (GSA) is utilized for synthesizing the CPMs [9–11]. The CPMs displayed on a spatial light modulator (SLM) generate an ensemble of sparse randomly distributed intensity dots on the sensor plane. The ensemble of dots obtained from the measured wavefront are cross-correlated with a reference dot pattern created by illuminating a single CPM with a plane wave. The intensity response in each cell of the array is shifted according to the average slope of the wavefront in each corresponding cell. Composing the entire slopes yields a three-dimensional curvature used as an approximation to the wavefront incident on the SHWS.

The optical scheme of COACH-based SHWS is shown in Fig. 1(a). A plane wave passes through a phase object and the emitted wavefront is sampled by an array of identical phase elements each of which is a product of the CPM and a lens transmittance. The lens performs a Fourier transform which according to the GSA is needed for the response of the dots on the camera plane. As in conventional SHWS, in front of each CPM, the local wavefront emitted from the phase object is approximated to a tilted plane wave. The goal of the system is to measure the tilt angle of each local planar wavefront and to estimate the global curvature by fusing all the local angles. As in ordinary SHWS, estimating the tilt angle is done by measuring the shift of the intensity response recorded on the camera. Since in COACH-based SHWS, this intensity response is the ensemble of dots, their common shift is measured by cross-correlation with a reference pattern. As much as the cross-correlation peak is sharper, the accuracy of the measurement is higher. Therefore, to narrow the peak as much as possible, the cross-correlation is done by a non-linear process with two parameters optimized to yield the sharpest cross-correlation peak.

We start the formal analysis by calculating the distribution of the reference pattern. Each cell of the size $a \times a$ in the CPM designed by the GSA has the distribution of $t(x,y)=\exp[i\varphi(x,y)]\text{Rect}(x/a,y/a)$, where $\varphi(x,y)$ is a chaotic phase function and $\text{Rect}(x/a,y/a)=1$ for all $|x|,|y|<a/2$ and 0 otherwise. $t(x,y)$ is attached to a diffractive microlens with $f$ focal length. It is well-known that illuminating these combined phase elements at a single cell of the array with a plane wave yields the following intensity [12],

$$I_R(x,y) = \left| V[1/\lambda f] \mathcal{F}\{t(x,y)\} \right|^2, \qquad (1)$$

where $\mathcal{F}$ is 2D Fourier transform and $\nu[\cdot]$ is the scaling operator such that $\nu[1/b]g(x)=g(bx)$. $I_R(x,y)$ plays the role of the reference pattern in the cross-correlation. As mentioned above $\varphi(x,y)$ is designed by the GSA to yield an ensemble of sparse randomly distributed dots as a result of a Fourier transform of $\exp[i\varphi(x,y)]$ [9]. Therefore,

$$I_R(x,y) = \left|\nu[1/\lambda f]\mathcal{F}\left\{\exp[i\varphi(x,y)]\mathrm{Rect}(x/a, y/a)\right\}\right|^2$$
$$\cong \sum_{p=1}^{P} C_p \mathrm{sinc}^2\left(a\frac{x-x_p}{\lambda f}, a\frac{y-y_p}{\lambda f}\right), \quad (2)$$

where $\mathrm{sinc}[a \cdot (x,y)] = \mathcal{F}\{\mathrm{Rect}(x/a, y/a)\}$, $C_p$ is appositive constant, $P$ is the number of dots, and $\{x_1,...,x_P\}$ is a set of random numbers determined by the GSA. $I_R(x,y)$ is measured once by illuminating the central CPM cell in the array with a plane wave propagating in the z-direction and without the presence of the phase object.

Once the reference function is known, the phase object is introduced into the setup and wavefront sensing is performed. The wavefront coming from the phase object is divided into $M \times N$ subareas as the number of phase elements at the array. At each subarea, the wavefront is approximated to a tilted planar wavefront with a tilt angle that should be measured to estimate the shape of the overall wavefront. Since each phase element in the array is multiplied by approximated linear phase, the dot response in the Fourier plane is shifted a distance proportional to the tilt angle. The intensity response at the $(m,n)$-th subarea on the camera for the tilted planar wavefront with a tilt angle of $(\theta_m, \theta_n)$ is,

$$I_{m,n}(x,y) = \left|\nu[1/\lambda f]\mathcal{F}\left\{\exp[(i2\pi/\lambda)(x\sin\theta_m + y\sin\theta_n)\right.\right.$$
$$\left.\left. +i\varphi(x,y)]\mathrm{Rect}(x/a, y/a)\right\}\right|^2 \quad (3)$$
$$\cong \sum_{p=1}^{P} C_p \mathrm{sinc}^2\left(a\frac{x-x_p-f\tan\theta_m}{\lambda f}, a\frac{y-y_p-f\tan\theta_n}{\lambda f}\right).$$

Cross-correlation between $I_{m,n}(x,y)$ and $I_R(x,y)$ yields a correlation peak at the distance $(f \cdot \tan\theta_m, f \cdot \tan\theta_n)$ from the center of the $(m,n)$-th subarea. Measuring this shift of the peak gives an estimation on the value of the tilt angle of $(\theta_m, \theta_n)$, and collecting all the local angles of the entire $M \times N$ subareas enables the restoration of the wavefront emitted from the phase object. The accuracy of the peak shift is dependent on the peak width, and therefore we propose to cross-correlate the signals in a nonlinear way with two parameters optimized to produce the narrowest peak [10]. To understand the relation of the peak width with the parameters of the nonlinear cross-correlation, we consider the Fourier transforms of the two correlated functions as follows,

$$\tilde{I}_{m,n}(u,v) = \mathcal{F}\{I_{m,n}(x,y)\} = |\tilde{I}_{m,n}(u,v)|$$
$$\times \exp[i\Phi(u,v) + i \cdot f(u\tan\theta_m + v\tan\theta_n)] \quad (4)$$
$$\tilde{I}_R(u,v) = \mathcal{F}\{I_R(x,y)\} = |\tilde{I}_R(u,v)|\exp[i\Phi_R(u,v)],$$

where $\Phi(u,v)$ and $\Phi_R(u,v)$ are the phase of $\tilde{I}_{m,n}(u,v)$ [for $(\theta_m, \theta_n)=(0,0)$] and $\tilde{I}_R(u,v)$, respectively. Since $|\tilde{I}_{m,n}(u,v)| = |\tilde{I}_R(u,v)|$ and $\Phi(u,v) = \Phi_R(u,v)$, the nonlinear cross-correlation (NCC) with the optimized parameters $\alpha$ and $\beta$ becomes [10],

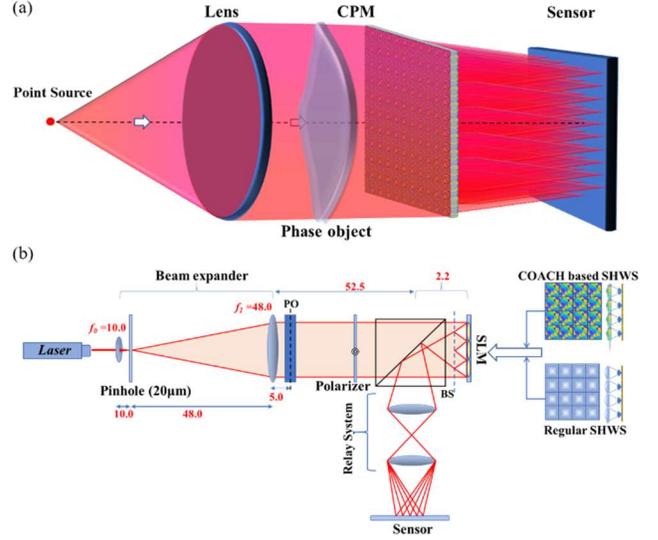

**Fig. 1.** (a) Optical Schematic and (b) experimental setup for the COACH-based SHWS. PO - Phase object, BS - Beamsplitter. All distances in (b) are in centimeters.

$$C(x,y) = \left|\mathcal{F}^{-1}\left\{|\tilde{I}_{m,n}(u,v)|^\alpha \exp[i\Phi(u,v)\right.\right.$$
$$\left.\left. +i \cdot f(u\tan\theta_m + v\tan\theta_n)]|\tilde{I}_R(u,v)|^\beta \exp[-i\Phi_R(u,v)]\right\}\right| \quad (5)$$
$$= \left|\tilde{I}_R(u,v)\right|^{\alpha+\beta}\exp[i \cdot f(u\tan\theta_m + v\tan\theta_n)]$$
$$= \Lambda(x - f\tan\theta_m, y - f\tan\theta_n).$$

$\Lambda(\cdot)$ is the peak function which should be as narrow as possible to increase the accuracy of the tilt angle measurement. Theoretically, the values of $\alpha$ and $\beta$ that yield the sharpest peak are the values that satisfy the equation $\alpha + \beta = 0$. However, the noisy experimental environment usually yields different optimal parameters and may not follow this equation. In this study, the optimal parameters $\alpha$ and $\beta$ of the NCC are found by minimizing the mean square error (MSE), because minimum MSE is the measure of accuracy which we aim to increase. The MSE assesses the reconstruction error by calculating the deviation of the reconstructed wavefront from the digitally simulated wavefront. The MSE is calculated as,

$$\mathrm{MSE} = \frac{1}{L \cdot K}\sum_{l=1}^{L}\sum_{k=1}^{K}\left(\hat{W}_{k,l} - W_{k,l}\right)^2, \quad (6)$$

where $\hat{W}_{kl}$ and $W_{kl}$ are the digital and experimental reconstructed wavefronts, respectively. $K$, $L$ are the numbers of data points in the wavefront. In case the measured wavefront is a-priori unknown, two strategies can be followed. First, the values of $\alpha$ and $\beta$ can be calibrated with a phase object that its wavefront is a-priori known and then the same $\alpha$ and $\beta$ can be used for the unknown wavefront. Alternatively, $\alpha$ and $\beta$ can be found directly for the unknown wavefront, using an optimization function that does not need any reference wavefront [10,11,13]. Following the procedure of the NCC, the center of the correlation peak for each cell is calculated by the center-of-mass method. The average slope of the tested wavefront in each cell is directly related to the displacement of the correlation peak from the center of the cell. Once the slope for each sub-aperture is calculated, then with the help of the zonal reconstruction technique [14,15], these slope values are fused to reconstruct the full wavefront.

Based on our previous investigations [8,9,11], it is well-known that cross-correlation between two bipolar functions reduces the noise and increases the sharpness of the correlation peaks. Therefore, to further increase the accuracy of the tilt angle measurement, we tested the option of using bipolar functions by recording two camera shots with two independent CPMs and with two different dot responses, where one response is subtracted from the other. In this scheme, the bipolar reference pattern is,

$$I_R(x,y) = \left|\nu[1/\lambda f]\mathcal{F}\{t_1(x,y)\}\right|^2 - \left|\nu[1/\lambda f]\mathcal{F}\{t_2(x,y)\}\right|^2, \quad (7)$$

where $t_j(x,y)=\exp[i\varphi_j(x,y)]\text{Rect}(x/a,y/a)$, and $\varphi_j(x,y)$ is the phase distribution of the $j$-th CPM ($j$=1,2). $I_{m,n}(x,y)$ is also obtained by two exposures with the same two CPMs so that its distribution is,

$$\begin{aligned}I_{m,n}(x,y) &= \left|\nu\left[\frac{1}{\lambda f}\right]\mathcal{F}\left\{\exp\left[\frac{i2\pi}{\lambda}(x\sin\theta_m+y\sin\theta_n)\right]t_1(x,y)\right\}\right|^2 \\ &- \left|\nu\left[\frac{1}{\lambda f}\right]\mathcal{F}\left\{\exp\left[\frac{i2\pi}{\lambda}(x\sin\theta_m+y\sin\theta_n)\right]t_2(x,y)\right\}\right|^2 \\ &\cong \sum_{p=1}^P C_{1,p}\text{sinc}^2\left(\frac{x-x_{1,p}-f\tan\theta_m}{\lambda f/a},\frac{x-x_{1,p}-f\tan\theta_m}{\lambda f/a}\right) \\ &- \sum_{p=1}^P C_{2,p}\text{sinc}^2\left(\frac{x-x_{2,p}-f\tan\theta_m}{\lambda f/a},\frac{y-y_{2,p}-f\tan\theta_n}{\lambda f/a}\right),\end{aligned} \quad (8)$$

where the two random series $\{x_{1,1},...x_{1,P}\}$ and $\{x_{2,1},...x_{2,P}\}$ are independent. The rest of the process is identical to the above-mentioned description of the process of the unipolar patterns.

The schematic of the experimental setup for COACH-based SHWS is shown in Fig. 1(b). A HeNe laser with a beam size of 8mm is used as the source, where a pair of lenses ($f_1$=10cm; $f_2$=48cm) is used to expand the beam diameter from 8mm to about 38mm. A pinhole of 20μm diameter used as a lowpass filter is placed at the common focal plane of the two lenses. The lowpass filter makes the wavefront after the beam expander as close as possible to a plane wave. This plane wave propagates through a phase object such that its phase distribution is represented by the wavefront emitted from the object. In the experiment, a positive lens of 30cm focal length is used as the phase object. The beam is polarized to the active orientation of the SLM (Holoeye PLUTO, 1920×1080 pixels, 8μm pixel pitch). The array of coded phase patterns displayed on the SLM is obtained by modulo-2π phase addition of the CPMs with the lenslet array of $f$=6.5mm focal length. The reflective SLM and the focal length enforce using a beamsplitter to reflect the modulated light coming from the SLM toward a digital camera (PCO.Edge 5.5 CMOS, pixel pitch=6.5μm, 2560×2160 pixel). An optical relay system is used to translate the focal plane of the microlens array onto the sensor plane. For comparison with our proposed method, we measured the wavefront with a regular SHWS implemented on the same setup by changing the phase pattern on the SLM to a lenslet array of $f$=6.5mm focal length. The central 1020×1020 pixels of the SLM were used to display an array of 30×30 CPMs attached to microlenses. The size of each cell is 34×34 pixels. The experimental results were compared with the digital wavefront. The digital wavefront is created by numerical Fresnel propagation from the phase object to the SLM plane using MATLAB with the same parameters as the experimental setup. The MSE was calculated between the digital wavefront and the tested wavefront reconstructed by the regular and by the COACH-based SHWS systems. For COACH-based SHWS, both techniques, unipolar and bipolar, were tested. To find the optimal number of dots of the sparse response, numbers of 3, 5, 10, and 12 dots were tested.

Figures 2($a_1$) and 2($a_2$) present the intensity response for two different CPMs when only the central cell is activated in the array. The dot structure is clearly seen in both figures and in the bipolar pattern of Fig. 2($a_3$) obtained as the difference of the intensities of Figs. 2($a_1$) and 2($a_2$). Figures 2($b_1$) and 2($b_2$) show the intensity response in the center of the COACH-based array equipped with two different CPMs, where the tested wavefront is introduced into the system. The corresponding bipolar response is shown in Fig. 2($b_3$). Figure 3 shows the experimental results of the NCC in comparison with the focal spot of a single microlens shown in Fig. 3($a$). The autocorrelation of Fig. 2($a_1$) by the NCC technique with ($\alpha$=0.0; $\beta$=1.0) and ($\alpha$=0.1; $\beta$=0.9) is shown in Figs. 3(b) and 3(c), respectively. The autocorrelation of Fig. 2($a_3$) by NCC technique with ($\alpha$=0.0; $\beta$=1.0) and ($\alpha$=0.1; $\beta$=0.9) is shown in Figs. 3(d) and 3(e), respectively. The NCC parameters (i.e. $\alpha$ and $\beta$) can vary between -1 to +1 and are optimized by searching the minimum MSE between the experimental and the digital wavefronts. With the parameters ($\alpha$=0.0; $\beta$=1.0) and ($\alpha$=0.1; $\beta$=0.9), the wavefront reconstruction result was the best with the lowest MSE in comparison to other values. Figure 3(f) shows the intensity response of regular SHWS for the test wave. Figure 3(h) shows the NCC of Fig. 2($a_3$) with Fig. 2($b_3$) [when the scale of 2($a_3$) is the same as one cell of 2($b_3$)] with the parameters $\alpha$=0.0 and $\beta$=1.0. The cross-section plots of correlation peaks with optimized values of $\alpha$ and $\beta$ are shown in Fig. 3(g) in comparison to the curve of the microlens focal spot (blue). Based on the plots of Fig. 3(g), it is clear that the NCC technique gives a sharper and narrower peak in comparison with the microlens focal spot of the regular SHWS.

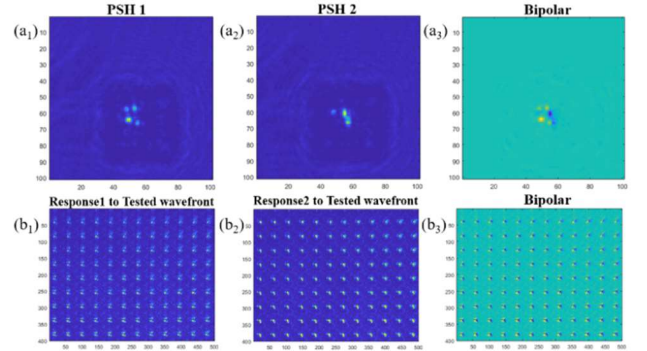

**Fig. 2.** A single intensity response recoded by the sensor for ($a_1$) CPM1 and ($a_2$) CPM2 displayed on the SLM. The bipolar pattern in ($a_3$) is obtained as the difference between ($a_1$) and ($a_2$); ($b_1$, $b_2$) The central part of two intensity patterns on the sensor for the two CPM arrays and for the input of the test wavefront; ($b_3$) the corresponding bipolar pattern obtained as the difference between ($b_1$) and ($b_2$).

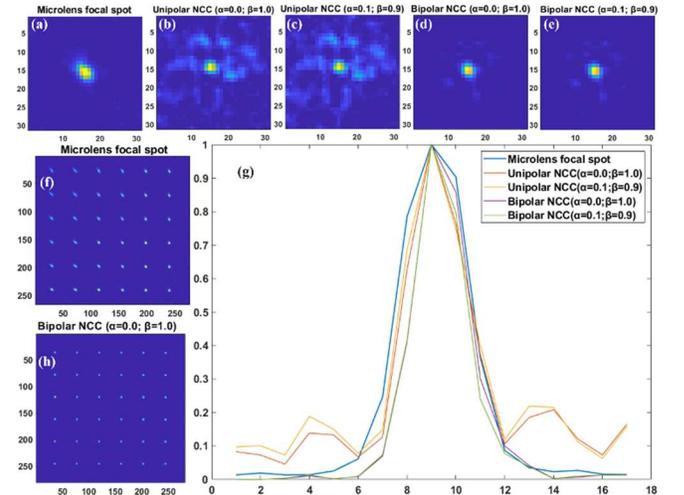

**Fig. 3.** (a) intensity response of a single microlens focal spot. Correlation peaks by unipolar NCC with (b) $\alpha$=0.0, $\beta$=1.0, (c) $\alpha$=0.1, $\beta$=0.9, and by bipolar NCC with (d) $\alpha$=0.0, $\beta$=1.0, (e) $\alpha$=0.1, $\beta$=0.9. (f) intensity response of microlens array for the test wavefront, (g) Horizontal cross-section of the microlens focal spot, unipolar and bipolar correlation peaks. (h) correlation peaks with bipolar NCC ($\alpha$=0.0, $\beta$=1.0) for the test wavefront.

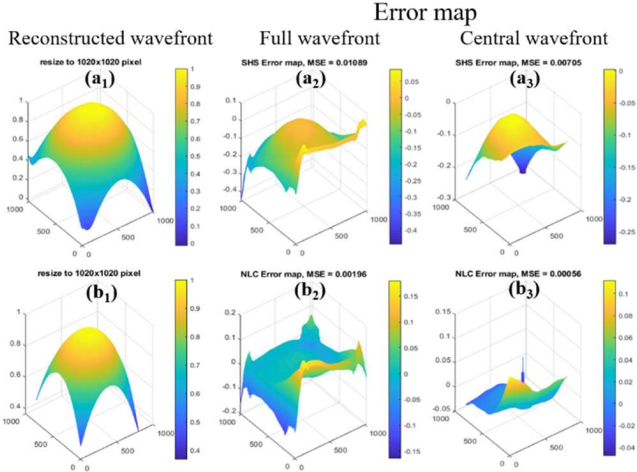

**Fig. 4.** wavefront reconstruction results with regular SHWS, and bipolar coded aperture SHWS with 5-sparse dot pattern. ($a_1$) is the resized reconstructed wavefront by the regular SHWS method. ($a_2$) is the error map of the wavefront in comparison with the digital wavefront. ($a_3$) is the error map when only the central part of both wavefronts is considered. ($b_1$, $b_2$, $b_3$) are the same reconstruction results as ($a_1$, $a_2$, $a_3$) for the bipolar COACH-based SHWS.

The reconstruction results of both the unipolar and bipolar SHWS are compared with the regular SHWS in Fig. 4. For the experiments, 1020×1020 pixels of SLM were used to display a 30×30 array of microlenses and CPMs. The 30×30 data points of a reconstructed wavefront are resized into 1020×1020 data points and compare with the simulated digital wavefront. Figures 4($a_1$) and 4($b_1$) show the 3D maps of the reconstructed wavefronts by the regular SHWS and bipolar COACH-based SHWS with 5 sparse dots, respectively. The 3D error maps between the reconstructed and digital wavefront for both the cases are shown in Figs. 4($a_2$) and 4($b_2$). The calculated MSE values of the various tested SHWS for each number of sparse dots are given in Table 1. From the error maps in Figs. 4($a_2$) and 4($b_2$), it can be seen that the boundary region of the wavefront extremely deviated from the digital shape resulting in higher MSE values because of the peripheral areas. Therefore, the 3D error maps for the centrally cropped 800×800 data points for both cases are shown in Figs. 4($a_3$) and 4($b_3$), respectively. From the flat error map in Fig. 4($b_3$), it is clear that the reconstructed wavefront of the bipolar method has a closer fit to the digital wavefront. MSE values of the centrally cropped wavefront are also calculated and given in Table 2. The MSE of 5-dot patterns for the unipolar case and most of the 3-dot patterns for the bipolar method yield minimum MSE among the tested cases and these minimum values are bolded in the tables. The case of 3-dot bipolar COACH-based SHWS has a smaller MSE by order of magnitude in comparison to the regular SHWS. The lower MSE values confirm the improvement in the wavefront reconstruction accuracy of the COACH-based SHWS. Furthermore, it should be noted that all the tested bipolar reconstruction results are better than the 5-dot bipolar but still the 5-dot unipolar gives better reconstruction results in comparison to the regular SHWS and to unipolar cases with other dot numbers.

In conclusion, we have presented a new SHWS technique with improved measurement accuracy. Coded phase masks are synthesized using the modified GSA algorithm to produce sparse dot patterns. Further, the bipolar sparse dot patterns are generated to reduce background noise. The experimental results with a different number of dot patterns are compared with the regular SHWS and the results confirm the validity and effectiveness of our method. From the calculated MSE values it can be noted that for the best case of 3-dot bipolar (overall 6 dots), the accuracy of COACH-based SHWS is improved by one order of magnitude in comparison with the regular SHWS. The cost of such improvement is the need for two camera shots. In case the speed of the measurement is more critical than the accuracy, the single-shot unipolar COACH-based SHWS can offer approximately double the accuracy than of the regular SHWS.

**Table 1. MSE values of SHWS for different number of dots**

| No. of Sparse Dots | Regular SHWS | Unipolar ($\alpha$=0.0; $\beta$=1.0) | Unipolar ($\alpha$=0.1; $\beta$=0.9) | Bipolar ($\alpha$=0.0; $\beta$=1.0) | Bipolar ($\alpha$=0.1; $\beta$=0.9) |
|---|---|---|---|---|---|
| 3 | 0.01089 | 0.10920 | 0.07920 | **0.00075** | **0.00150** |
| 5 | 0.01089 | **0.00944** | **0.00889** | 0.00181 | 0.00196 |
| 10 | 0.01089 | 0.08654 | 0.05396 | 0.00563 | 0.00313 |
| 12 | 0.01089 | 0.06623 | 0.04522 | 0.00528 | 0.00612 |

**Table 2. MSE value of centrally cropped wavefront**

| No. of Sparse Dots | Regular SHWS | Unipolar ($\alpha$=0.0; $\beta$=1.0) | Unipolar ($\alpha$=0.1; $\beta$=0.9) | Bipolar ($\alpha$=0.0; $\beta$=1.0) | Bipolar ($\alpha$=0.1; $\beta$=0.9) |
|---|---|---|---|---|---|
| 3 | 0.00705 | 0.12840 | 0.08820 | **0.00045** | 0.00130 |
| 5 | 0.00705 | **0.00379** | **0.00378** | 0.00087 | **0.00056** |
| 10 | 0.00705 | 0.08169 | 0.05426 | 0.00310 | 0.00192 |
| 12 | 0.00705 | 0.05649 | 0.04280 | 0.00246 | 0.00165 |

**References**

1. S. Goelz, J. J. Persoff, G. D. Bittner, J. Liang, C.-F. T. Hsueh, and J. F. Bille, Active." New wavefront sensor for metrology of spherical surfaces," Proc. SPIE 1542, Active and Adaptive Optical Systems, (1 December 1991)
2. D. Dayton, J. Gonglewski, B. Pierson, and B. Spielbusch, "Atmospheric structure function measurements with a Shack–Hartmann wave-front sensor," Optics Letters **17**, 1737-1739 (1992).
3. X. Tao, B. Fernandez, O. Azucena, M. Fu, D. Garcia, Y. Zuo, D. C. Chen, and J. Kubby, "Adaptive optics confocal microscopy using direct wavefront sensing," Optics Letters **36**, 1062-1064 (2011).
4. Y. Zhang, C. A. Girkin, J. L. Duncan, and A. Roorda, "Adaptive optics scanning laser ophthalmoscopy," Advanced Biophotonics: Tissue Optical Sectioning **10**, 507-557 (2013).
5. S.-W. Bahk, P. Rousseau, T. A. Planchon, V. Chvykov, G. Kalintchenko, A. Maksimchuk, G. A. Mourou, and V. Yanovsky, "Generation and characterization of the highest laser intensities ($10^{22}$ W/cm$^2$)," Optics Letters **29**, 2837-2839 (2004).
6. A. Talmi and E. N. Ribak, "Wavefront reconstruction from its gradients" Journal of the Optical Society of America A **23**, 288-297 (2006).
7. A. Vijayakumar and J. Rosen, "Interferenceless coded aperture correlation holography–a new technique for recording incoherent digital holograms without two-wave interference" Optics Express **25**, 13883-13886 (2017).
8. J. Rosen, A. Vijayakumar, M. R. Rai, S. Mukherjee, and A. Bulbul "Review of 3D Imaging by Coded Aperture Correlation Holography (COACH)," Appl. Sci. **9**(3), 605 (2019).
9. M. R. Rai and J. Rosen, "Noise suppression by controlling the sparsity of the point spread function in interferenceless coded aperture correlation holography (I-COACH)," Optics Express **27**, 24311-24323 (2019).
10. M. R. Rai, A. Vijayakumar, and J. Rosen, "Non-linear adaptive three-dimensional imaging with interferenceless coded aperture correlation holography (I-COACH),"Optics Express **26**, 18143 (2018).
11. N. Dubey, J. Rosen, and I. Gannot, "High-resolution imaging system with an annular aperture of coded phase masks for endoscopic applications," Optics Express **28**, 15122-15137 (2020).
12. J. W. Goodman, "Introduction to Fourier Optics," Roberts & Co. (2005).

13. C. Liu, T. Man, and Y. Wan, "Optimized reconstruction with noise suppression for interferenceless coded aperture correlation holography," Appl. Opt. **59**, 1769-1774 (2020).
14. G. Dai, "Wavefront Optics for Vision Correction," (SPIE Press, 2008, n.d.).
15. W. H. Southwell, "Wave-front estimation from wave-front slope measurements" Journal of the Optical Society of America **70**, 998-1006 (1980).